\documentstyle[preprint,eqsecnum,aps,amsmath,amsfonts,amssymb]{revtex}
\def\mR{{\mathbb{R}}}
\def\mS{{\mathbb{S}}}
\def\mZ{{\mathbb{Z}}}

\def\displayfrac#1#2{\frac{\displaystyle #1}{\displaystyle #2}}

\begin{document}
\title{Wigner-Weyl isomorphism for quantum mechanics on Lie groups}
\author{N. Mukunda\thanks{email: nmukunda@cts.iisc.ernet.in}}
\address{Centre for Theoretical Studies, Indian Institute of Science,
Bangalore~560~012, India}
\author{G. Marmo\thanks{email: Giuseppe.Marmo@na.infn.it}}
\address{Dipartimento di Scienze Fisiche, Universita di Napoli Federico II and
INFN, Via Cinzia, 80126 Napoli, Italy}
\author{Alessandro Zampini\thanks{email:Alessandro.Zampini@na.infn.it}}
\address{Dipartimento di Scienze Fisiche, Universita di Napoli Federico II and
INFN, Via Cinzia, 80126 Napoli, Italy}
\author{S. Chaturvedi\thanks{e-mail: scsp@uohyd.ernet.in}}
\address{ School of Physics, University of Hyderabad, Hyderabad 500 046,
India}
\author{ R.Simon\thanks{email: simon@imsc.res.in}}
\address{The Institute of
 Mathematical Sciences, C. I. T. Campus, Chennai 600 113, India}
 \date{\today}
\maketitle
\begin{abstract}
The Wigner-Weyl isomorphism for quantum mechanics on a compact simple
Lie group $G$ is developed in detail. Several New features are shown to arise 
which have no counterparts in the familiar Cartesian case. Notable among 
these is the notion of a `semiquantised phase space', a structure on which 
the Weyl symbols of operators turn out to be naturally defined and, 
figuratively speaking, located midway between the classical phase space $T^*G$
and the Hilbert space of square integrable functions on $G$. 
General expressions for the star product for Weyl symbols are presented and 
explicitly worked out for the angle-angular momentum case. 
\end{abstract}
\section{Introduction}
It is well known that the method of Wigner distributions \cite{1}, which
describes every state of a quantum mechanical system by a
corresponding real quasi probability density on the classical
phase space, is dual to the Weyl mapping \cite{2} of classical dynamical
variables to quantum mechanical operators. Together they provide
the Wigner-Weyl isomorphism, whereby both states and operators in
quantum mechanics can be given c-number descriptions on the
classical phase space. The trace of the product of two operators
is then calculable as the integral of the product of the two
corresponding `Weyl symbols' or phase space functions. Combined
with the work of Moyal\cite{3}, which shows how products and commutators
of operators are expressed in phase space language, this entire
development may be called the Wigner-Weyl-Moyal or WWM method in
quantum mechanics and has been instrumental in giving rise to the 
fertile subject of deformation quantisation\cite{4}. 
An important feature of the Wigner distribution
is that while it is not by itself a phase space probability
density, its marginals obtained by respectively integrating over
momenta or over coordinates do reproduce the quantum mechanical
expressions for probability densities in coordinate and in
momentum space respectively.

The WWM method has been studied most extensively in the case of
Cartesian systems in quantum mechanics. By this we mean those
systems whose configuration space  Q is $\mR^n$ for some integer
$n~\geq~1$. The classical phase space is then $T^* Q \simeq
\mR^{2n}$. While Schr\"odinger wave functions are square integrable
functions on $\mR^n$ , both Wigner distributions and Weyl symbols
are functions on $\mR^{2n}$. Quantum kinematics can be expressed via
the Heisenberg canonical commutation relations for Cartesian
coordinates and their conjugate momenta, or via the exponentiated
Weyl form using families of unitary operators. An important
feature in this case is that as far as their eigenvalue spectra
are concerned, the momenta do not experience any quantisation on
their own; they account for the second factor in $T^* Q \simeq
\mR^{2n}\simeq \mR^n \times \mR^n$. Furthermore we have in this case the
Stone-von Neumann theorem on the uniqueness of the irreducible
representation of the Heisenberg commutation relations, and the
important roles of the groups $Sp(2n,R)$ and $Mp(2n)$
corresponding to linear canonical transformations on coordinates
and momenta.

There has been for some time considerable interest in developing
the Wigner-Weyl isomorphism for other kinds of quantum systems,
that is, for non Cartesian systems\cite{5}-\cite{16}. 
In these cases, typically the
underlying quantum kinematics cannot be expressed by
Heisenberg-type commutation relations. The situations studied
include the quantum mechanics of an angle-angular momentum pair,
where the configuration space is $Q=\mS^1$ \cite{17},\cite{18}, and
finite state quantum systems corresponding to a finite dimensional
Hilbert space \cite{19},\cite{20}. More recently,  the method of
Wigner distributions has been developed for quantum systems whose
configuration space is a compact simple Lie group ; and in the
discrete case when it is a finite group of odd order \cite{21}
\cite{22}. In all these departures from the Cartesian situation, an
important aspect is the occurrence of new features which do not
show up at all with Cartesian variables.

The aim of the present work is to develop in detail the
Wigner-Weyl isomorphism for quantum mechanics on a compact simple
Lie group. Here the configuration space $Q$ is a (compact simple)
Lie group $G$, so the corresponding classical phase space is
$T^*G\simeq G\times \underline{G}^*$, where $\underline{G}^*$ is
the dual to the Lie algebra $\underline{G}$ of $G$. In the quantum
situation, Schr\"odinger wave functions are complex square
integrable functions on $G$, and observables or dynamical
variables are linear hermitian operators acting on such functions.
The replacements for the canonical Heisenberg commutation
relations are best formulated using the (commutative) algebra of
suitable smooth functions on $G$, and (say) the left regular
representation of $G$ acting on functions on itself. The natural
question that arises in trying to set up a Wigner-Weyl isomorphism
in this case is whether quantum states and operators are to be
described using suitable functions on the classical phase space
$T^*G$. In \cite{21} an over complete Wigner distribution formalism
for quantum states, which transforms in a reasonable way under
left and right group actions and also reproduces the natural
marginal probability distributions, has been developed. The
methods developed there are here exploited to set up a Wigner-Weyl
isomorphism in full detail, disclosing many interesting
differences compared to the Cartesian case. In particular we find
that this isomorphism does not directly utilise c-number functions
on $T^*G$ at all, but instead uses a combination of functions on
$G$ and operators on a simpler Hilbert space, standing in a sense
midway between $T^*G$ and the Hilbert space of the quantum system.
This feature is traceable to the non Abelian nature of $G$,
something which is absent in the Cartesian case when $Q$ is the
Abelian group $\mR^n$.

 The material of this paper is organised as follows. In Section II we 
briefly recapitulate key features of the Wigner-Weyl isomorphism for the
Cartesian and angle-angular momentum cases. This sets the stage for Section
III where we develop the quantum kinematics for situations where the 
configuration space is a compact Lie group and thus go beyond the Abelian 
cases discussed in Section I. This analysis leads to a proper identification 
of the analogues of the `momenta' of the Cartesian case and helps set up the 
Wigner distribution for such situations possessing properties expected of a 
Wigner distribution. The Wigner distributions so defined have a certain degree
of overcompleteness about them, a circumstance forced by the non Abelian
nature of the underlying group $G$. A key ingredient in this construction is 
the notion of the `mid point' of two group elements introduced in an earlier 
work \cite{21}. In Section IV  a more compact description in terms of 
Weyl symbols devoid of any redundances is developed and correspondences 
facilitating transition from the Cartesian case to more general situations are 
established. The results of Section IV are exploited in Section V towards 
defining a star product between Weyl symbols for operators and the general
expression for the star product is  explicitly worked out for the non
Cartesian, albeit Abelian case of angle-angular momentum. Section VI is
devoted to analysing the minimal structure on which the Weyl symbols for 
operators find their natural definition. This leads to the concept of a `non
commutative cotangent space'  or a `semiquantised phase space' the
ramifications  of which are examined further towards highlighting  
the structural similarity between classical phase space functions and the 
Weyl symbols. A short appendix contains some technical details concerning 
results used in Section V.    
    
 \section{The Wigner-Weyl isomorphism : Cartesian and angle-angular
momentum cases} In this Section we recall briefly the relevant
structures needed to set up the Wigner-Weyl isomorphism for
Cartesian quantum mechanics. This is to facilitate comparison with
the Lie group case later on. For simplicity we choose one degree
of freedom only, as the extension to $Q=\mR^n$ is straight forward.
We also recall the angle-angular momentum case, $Q=\mS^1$, where we
already see significant differences from the Cartesian case; these
increase when we go to $Q=G$.

\noindent \underline{One~ dimensional~ Cartesian~ quantum~
mechanics}

The canonical Heisenberg commutation relation between hermitian
coordinate and momentum operators ${\hat q}$ and ${\hat p}$,
fixing the kinematics, is
\begin{equation} [\hat{q},\hat{p}] = i.
 \end{equation}

 In the unitary Weyl form this is expressed as
follows: \begin{eqnarray}
U(p) = \exp(i\;p\;\hat{q})&&, V(q)= \exp (-i\;q\;\hat{p}):\nonumber\\
U(p)~V(q)&=& V(q)~U(p)e^{iqp},~~~q,~p\in \mR .
\end{eqnarray}

In the Cartesian case the exponentials can be combined to define a
phase space displacement operator
\begin{eqnarray}
D(q,p) &=& U(p)~V(q)e^{-iqp/2}=V(q)~U(p)e^{iqp/2}\nonumber\\
&=&\exp(i\;p\;\hat{q}-i\;q\;\hat{p}).
\end{eqnarray}

However this cannot be done even in the single angle-angular
momentum pair case, and also when we treat the Lie group case. We
therefore use expressions in which the exponentials are kept
separate.

The standard form of the unique irreducible representation of eqs.
(2.1,2.2) uses the Hilbert space of square integrable functions
$\psi(q)$ on $\mR$. Introducing as usual an ideal basis of
eigenvectors of ${\hat q}$ we have:
\begin{eqnarray}
{\cal H}=L^2(\mR)&=&\{\psi(q)~|~\parallel\psi\parallel^2=
\int\limits_{\mR} dq |\psi(q)|^2~<~\infty \},\nonumber\\
\psi(q) &=& \langle q|\psi\rangle,~{\hat q}|q>=q|q>,\nonumber\\
\langle q^{\prime}|q\rangle&=& \delta\left(q^{\prime}-q\right).
\end{eqnarray}

On such $\psi(q)$  (subject to relevant domain conditions) the
actions of ${\hat q},{\hat p}, U(p), V(q)$ are
\begin{eqnarray}({\hat q}~\psi)(q)&=&q~\psi(q),~~({\hat
p}~\psi)(q)=-i\frac{d}
{dq}\psi(q);\nonumber\\
(U(p^\prime)~\psi)(q)&=& e^{ip^\prime q}\psi(q),~~
(V(q^\prime)~\psi)(q)=\psi(q-q^\prime).
\end{eqnarray}

The momentum space description of $|\psi>$ uses the Fourier
transform of $\psi(q)$; in terms of the ideal eigenstates $|p>$
for ${\hat p}$:
\begin{eqnarray} \tilde{\psi}(p) &=& \langle p|\psi\rangle = \int\limits_{\mR}
\displayfrac{dq}{\sqrt{2\pi}}
\exp(-iqp)\psi(q),\nonumber\\
\parallel \psi\parallel^2 &=&
\int\limits_{\mR} dp |\tilde{\psi}(p)|^2.
\end{eqnarray}

 The displacement operators (2.2) form a complete trace orthonormal
set (in the continuum sense) in the space of operators on ${\cal
H}$:
\begin{eqnarray} \mbox{Tr}((U(p^\prime)V(q^\prime))^\dagger
U(p)V(q)) =2\pi\delta(q^\prime-q)~\delta(p^\prime-p).
\end{eqnarray}

The completeness property will be used later.

The definitions of the Wigner distribution for a normalised pure
state $|\psi> \in {\cal H}$, or more generally for a mixed state
with density operator ${\hat \rho}$, are
\begin{eqnarray}
W(q,p)&=&\frac{1}{2\pi} \int\limits_{\mR} dq^{\prime}
\psi\left(q-\frac{1}{2} q^{\prime}\right) \psi\left(q+\frac{1}{2}
q^{\prime}\right)^*
\exp\left(i\;p\; q^{\prime} \right),\nonumber\\
W(q,p)&=&\frac{1}{2\pi}\int\limits_{\mR} dq^{\prime} \langle
q-\frac{1}{2} q^{\prime}\big| \hat{\rho}\big| q +\frac{1}{2}
q^{\prime} \rangle \exp\left(i\;p\;q^{\prime}\right) ,
\end{eqnarray}

\noindent
(The dependences on $|\psi>,~{\hat \rho}$ are left
implicit). While $W(q,p)$ is real though not always non negative,
the recovery of the marginal position and momentum space
probability densities is assured by
\begin{eqnarray}
 \int\limits_{\mR} d p\; W(q,p) = \left\langle q \big|
\hat{\rho}\big| q \right\rangle , ~~~~ \int\limits_{\mR} dq\; W(q,p)
=\left\langle p \big| \hat{\rho}\big| p \right\rangle.
 \end{eqnarray}

  It is possible to express $W(q,p)$ in a more compact form by introducing
a family of hermitian operators ${\hat W}(q,p)$ on ${\cal H}$ with
interesting algebraic properties. They are essentially the double
Fourier transforms of the displacement operators (2.2):
\begin{eqnarray}
W(q,p)&=& \mbox{Tr}(\hat{\rho} {\hat W}(q,p)),\nonumber\\
{\hat W}(q,p)&=&{\hat W}(q,p)^\dagger=\frac{1}{(2\pi)^2}\int_{\mR}
\int_{\mR} dq^\prime~dp^\prime~U(p^\prime)~V(q^\prime)
e^{ipq^{\prime}-ip^{\prime}\left(q+ \frac{1}{2}q^{\prime}\right)}.
\end{eqnarray}

 It has been shown in \cite{17} that, apart from sharing the trace
orthonormality property (2.7) which is preserved by the Fourier
transformation,
\begin{eqnarray}
\mbox {Tr}({\hat W}(q^\prime,p^\prime){\hat W}(q,p))
=\frac{1}{2\pi}\delta(q^\prime-q)~\delta(p^\prime-p),
\end{eqnarray}

\noindent
we have the following behaviours under anticommutation
 with ${\hat q}$ and ${\hat p}$ :
 \begin{eqnarray}
\frac{1}{2}\{{\hat q},{\hat W}(q,p)\}=q\;
\hat{W}(q,p),~~\frac{1}{2}\{{\hat p},{\hat W}(q,p)\}=p
\;\hat{W}(q,p).
\end{eqnarray}

Thus we may regard ${\hat W}(q,p)$ as operator analogues of Dirac
delta functions concentrated at individual phase space points. In
\cite{19} they have been called `phase point operators'.

Turning to the Weyl map, it takes a general classical dynamical
variable, a (square integrable) function $a(q,p)$ on the classical
phase space, to a corresponding (Hilbert-Schmidt) operator ${\hat
A}$ on ${\cal H}$:
\begin{eqnarray}
 a(q,p)\rightarrow && {\tilde
a}(p^\prime,q^\prime) = \int_{\mR} \int_{\mR} dq ~dp\; a(q,p)~
e^{i(pq^\prime-qp^\prime)}\nonumber\\
\rightarrow && {\hat A}= \frac{1}{2\pi}\int_{\mR} \int_{\mR}
dq^\prime~dp^{\prime} \tilde{a}(p^{\prime},
q^{\prime})\;U(p^{\prime})~V(q^\prime) e^{-iq^\prime p^\prime/2}.
\end{eqnarray}

The important property of this map is that traces of operators on
${\cal H}$ go into integrals over phase space:
\begin{eqnarray}
\mbox{Tr}({\hat A}^\dagger~{\hat B})=\int_{\mR} \int_{\mR} dq ~dp
\;a(q,p)^*~ b(q,p).
\end{eqnarray}

\noindent One can immediately see that the relation between
$a(q,p)$ and ${\hat A}$ is given by
\begin{eqnarray}
{\hat A}=2\pi \int_{\mR} \int_{\mR} dq ~dp\; a(q,p)~{\hat W}(q,p),
\end{eqnarray}

\noindent thus establishing that the Wigner and Weyl maps are
inverses of one another. Indeed extending the definition of the
Wigner distribution (2.10) to a general operator ${\hat A}$ on
${\cal H}$, we have
\begin{eqnarray}
 a(q,p)=\mbox{Tr}({\hat A}~{\hat W}(q,p)).
\end{eqnarray}

\noindent It is this kind of isomorphism that we wish to develop
when $\mR$ is replaced by a compact simple Lie group $G$.

\noindent \underline{ The angle-angular momentum case}

We now trace the changes which appear if we replace the Cartesian
variable $q\in \mR$ by an angle $\theta \in (-\pi,\pi)$. The
corresponding hermitian operator is denoted by ${\hat \theta}$,
with eigenvalues $\theta$; its canonical conjugate ${\hat M}$ has
integer eigenvalues $m=0,\pm 1, \pm 2,\cdots$. Thus $m\in \mZ$
unlike the Cartesian $p$, so ${\hat M}$ is `already quantised'.
The replacements for eqs. (2.4,2.6) are
\begin{eqnarray}
 {\cal H}=L^2(\mS^1)&=&\{\psi(\theta)~|~\parallel\psi\parallel^2=
\int\limits_{-\pi}^{\pi} d\theta |\psi(\theta)|^2~<~\infty \},\nonumber\\
\psi(\theta) &=& \langle \theta|\psi\rangle,~\langle \theta^\prime
|\theta \rangle=\delta(\theta^\prime-\theta),~{\hat
\theta}|\theta>=\theta|\theta>,\nonumber\\
 \psi_m &=&\langle m|\psi\rangle=\frac{1}{\sqrt{2\pi}}
\int\limits_{-\pi}^{\pi} d\theta e^{-im\theta}\psi(\theta),\nonumber\\
~\parallel\psi\parallel^2 &=&\sum_{m\in \mZ} |\psi_m|^2;\nonumber\\
{\hat M}|m>&=&
m|m>,~~~~~~~<\theta|m>=\frac{1}{\sqrt{2\pi}}e^{im\theta}.
\end{eqnarray}

\noindent In place of the Heisenberg commutation relation (2.1),
we have only the exponentiated Weyl version:

\begin{eqnarray}
 U(m) = \exp(i\;m\;\hat{\theta})&&, V(\theta)= \exp
(-i\;\theta\;\hat{M}):
\nonumber\\
U(m)~V(\theta)&=& V(\theta)~U(m)e^{im\theta}.
\end{eqnarray}

With the actions
\begin{eqnarray}
(U(m^\prime)~\psi)(\theta)&=& e^{im^\prime \theta}\psi(\theta),\nonumber\\
(V(\theta^\prime)~\psi)(\theta)&=&\psi([\theta-\theta^\prime]),\nonumber\\
\left[\theta -\theta^\prime \right]&=&
\theta-\theta^\prime~~~\mbox{mod}~2\pi,
\end{eqnarray}

\noindent
we have an irreducible system on ${\cal H}=L^2(\mS^1)$.
 The analogues of the displacement operators (2.3)
 are now
\begin{eqnarray}
U(m)~V(\theta)e^{-im \theta/2} = V(\theta)~U(m)e^{im \theta/2},
\end{eqnarray}

\noindent
but here the exponents cannot be combined. They do
however form a complete trace orthonormal system :
\begin{eqnarray}
\mbox{Tr}((U(m^\prime)V(\theta^\prime))^\dagger U(m)V(\theta))
=2\pi\delta_{mm^\prime}~\delta(\theta^\prime-\theta).
\end{eqnarray}

With this preparation we can turn to the definition of the Wigner
distribution and the Weyl map. For a given density operator ${\hat
\rho}$ on ${\cal H}$, the former is
\begin{eqnarray}
W(\theta,m)&=&\frac{1}{2\pi}\int\limits_{-\pi}^{\pi}
d\theta^{\prime} \left\langle \theta-\frac{1}{2}
\theta^{\prime}\big| \hat{\rho}\big| \theta +\frac{1}{2}
\theta^{\prime} \right\rangle \exp\left(i\;m\;
\theta^{\prime}\right) .
\end{eqnarray}

\noindent
We see immediately that this is not a function on the
classical phase space $T^*\mS^1 \simeq \mS^1\times \mR$, which is a
cylinder, but on a `partially quantised' space $\mS^1\times \mZ$. We
may regard this space as standing somewhere in between $T^*\mS^1$
and the fully quantum mechanical  Hilbert space and operator set
up. The marginals are properly reproduced in the sense that
\begin{eqnarray}
\int\limits_{-\pi}^{\pi} d\theta\; W(\theta,m) &=&
\left\langle m \big| \hat{\rho}\big| m \right\rangle, \nonumber\\
 \sum_{m\in \mZ}~W(\theta,m) &=&\left\langle \theta \big|
\hat{\rho}\big| \theta \right\rangle.
\end{eqnarray}

We can display $W(\theta,m)$ as
\begin{eqnarray}
W(\theta,m)&=& \mbox{Tr}(\hat{\rho} {\hat W}(\theta,m)),\nonumber\\
{\hat W}(\theta,m)&=&{\hat W}(\theta,m)^\dagger=
\frac{1}{(2\pi)^2}\sum_{m^\prime \in \mZ} \int_{-\pi}^{\pi}
d\theta^\prime~~U(m^\prime)~V(\theta^\prime)
e^{im\theta^{\prime}-im^{\prime}\left(\theta+\frac{1}{2}
\theta^\prime\right)},
\end{eqnarray}

\noindent
and like their Cartesian counterparts these operators
form a trace orthonormal system:
\begin{eqnarray}
\mbox{Tr}({\hat W}(\theta^\prime,m^\prime){\hat W}(\theta,m))
=\frac{1}{2\pi}\delta(\theta^\prime-\theta)~\delta_{m m^\prime}.
\end{eqnarray}

In a similar spirit, the Weyl map now takes any `classical'
function $a(\theta, m)$ on $\mS^1\times \mZ$ into an operator on
$L^2(\mS^1)$:
\begin{eqnarray}
a(\theta,m)\rightarrow && {\tilde a}(m^{\prime},\theta^{\prime}) =
\sum_{m\in \mZ} \int_{-\pi}^{\pi} d\theta ~ a(\theta,m)~
e^{i(m \theta^\prime-m^\prime \theta)},\nonumber\\
\rightarrow && {\hat A}= \frac{1}{2\pi}\sum_{m^\prime \in \mZ}
\int_{-\pi}^{\pi} d\theta^{\prime}
~\tilde{a}(m^{\prime},\theta^{\prime})~U(m^{\prime})~V(\theta^{\prime})
e^{-im^{\prime} \theta^{\prime}/2}.
\end{eqnarray}

\noindent
Then the trace operation becomes, as in (2.14):
\begin{eqnarray}
\mbox{Tr}({\hat A}^\dagger~{\hat B})=\sum_{m\in \mZ}
\int_{-\pi}^{\pi}
 d\theta ~ a(\theta,m)^*~b(\theta,m).
\end{eqnarray}

\noindent Combining eqs.(2.24,2.26) we are able to get the
analogue to (2.15):
\begin{eqnarray}
{\hat A}=2\pi \sum_{m\in \mZ} \int_{-\pi}^{\pi} d\theta ~
a(\theta,m)~{\hat W}(\theta,m).
\end{eqnarray}

\noindent
In this way the similarities as well as important
differences compared to the Cartesian case are easily seen.

\section{Quantum kinematics in the Lie group case and the Wigner distribution}

Let $G$ be a (non Abelian) compact simple Lie group of order $n$,
with elements $g,g^\prime,\cdots$ and composition law $g^\prime,g
\rightarrow g^\prime g$. To set up the kinematics appropriate for
a quantum system with configuration space $Q=G$, it is simplest to
begin with the Hilbert space of `Schr\"odinger' wave functions.
The normalised left and right invariant volume element on $G$ is
written as $dg$. For suitable functions $f(g)$ on $G$ we have the
invariances and normalisation condition

\begin{eqnarray}
\int\limits_G dg\;f(g) &=& \int\limits_G dg\;
(f(g^{\prime}g)~\mbox{or}~f(g g^{\prime})~\mbox{or}~f(g^{-1})),
\nonumber\\
\int\limits_G dg &=& 1.
\end{eqnarray}

\noindent
Correspondingly we can introduce a Dirac delta function
on $G$ characterised by
\begin{eqnarray}
\int\limits_G dg\; (\delta(g^{\prime-1}g)~\mbox{or}~\delta(g
g^{\prime-1})~{\rm or}~ \delta(g^{-1}g^\prime)~\mbox
{or}~\delta(g^\prime g^{-1}))f(g)=f(g^\prime).
\end{eqnarray}

\noindent
Thus $\delta(g)$ is a delta function concentrated at the identity
element $e\;\in\;G$.

We take the Hilbert space ${\cal H}$ for the quantum system to be
made up of all complex square integrable functions on $G$:

\begin{eqnarray}
{\cal H} = L^2(G)=\left\{\psi(g) \in  {C}|
\parallel\psi\parallel^2 = \int\limits_G dg |\psi(g)|^2 <
\infty\right\}.
\end{eqnarray}

\noindent
A convenient basis of ideal vectors $|g>$ can be
introduced such that for a general $|\psi>\in {\cal H}$ we may
write
\begin{eqnarray}
\psi(g)=<g|\psi>,~~~<g^\prime|g>=\delta(g^\prime g^{-1}).
\end{eqnarray}

The notion of `position coordinates' is intrinsically captured by
the commutative algebra representing real valued smooth functions
$f(g)$ on $G$, i.e $f\in {\cal F}(G)$. To each such function we
associate a hermitian multiplicative operator ${\hat f}$ on ${\cal
H}$:
\begin{eqnarray}
f\in {\cal F}(G)\rightarrow {\hat f} &=& \int\limits_G dg\;f(g)
|g><g|,\nonumber\\
({\hat f}\psi)(g)&=&f(g)\psi(g).
\end{eqnarray}

\noindent Thus all these operators commute with one another, being
`diagonal' in the position description $\psi(g)$ of $|\psi>$.

To complete the kinematics and to obtain an irreducible system of
operators on ${\cal H}$ we have to adjoin suitable `momenta'. Here
we have two choices, corresponding to the left and right
translations of $G$ on itself by group action. We choose the
former, and so define a family of unitary operators $V(g)$ to give
the left regular representation of $G$ :
\begin{eqnarray}
(V(g^\prime )\psi)(g)&=&\psi(g^{\prime -1}g),\nonumber\\
V(g^\prime)|g>&=&|g^\prime g>.
\end{eqnarray}

\noindent
They obey
\begin{eqnarray}
V(g^\prime )V(g )&=&V(g^\prime g ), \nonumber\\
V(g)^\dagger V(g) &=& I.
\end{eqnarray}

\noindent To identify their hermitian generators, we introduce a
basis $\{e_r\}$ in the Lie algebra  $\underline{G}$ of $G$. Using
the exponential map $\underline{G} \rightarrow G$, we write a
general $g\in G$ as
\begin{eqnarray}
g=\exp(\alpha^r~e_r),
\end{eqnarray}

\noindent
the sum on $r$ being from $1$ to $n$. The generators
${\hat J}_r$ of $V(g)$ are then identified by
\begin{eqnarray}
V(\exp({\alpha}^r ~e_r))= \exp(-i{\alpha}^r~{\hat J}_r).
\end{eqnarray}

\noindent
These are hermitian operators on the Hilbert space ${\cal
H}$, obeying commutation relations involving the structure
constants $C_{rs}^t$ of $G$:
\begin{eqnarray}
\left[\hat{J}_r, \hat{J}_s\right] &=& i\;C_{rs}\;^t \hat{J}_t.
\end{eqnarray}

\noindent
On Schr\"odinger wave functions $\psi(g)$ each ${\hat
J}_r$ acts as a first order partial differential operator; indeed
if the (right invariant) vector fields generating the left action
of $G$ on itself are written as $X_r$, then we have
\begin{eqnarray}
{\hat J}_r~\psi(g)=i~X_r~\psi(g).
\end{eqnarray}

\noindent The commutation relations (3.10) are direct consequences
of similar commutation relations among the vector fields $X_r$.

The analogue of the Cartesian Heisenberg-Weyl system (2.1,2.2) is
now obtained by putting together the following ingredients:
\begin{eqnarray}
f_1,~f_2\in {\cal F}(G)\rightarrow &&{\hat f}_1{\hat f}_2=
{\hat f}_2{\hat f}_1;\nonumber\\
f \in {\cal F}(G),~g^\prime \in G \rightarrow &&V(g^\prime)~{\hat
f}~V(g^\prime)^{-1}={\hat f}^\prime,\nonumber\\
&&f^\prime(g)=f(g^{\prime -1}g);
\end{eqnarray}

\noindent along with the representation property (3.7) for $V(g)$.
This is in the spirit of the unitary Weyl system (2.2). In
infinitesimal terms we have
\begin{eqnarray}
\left[\hat{J}_r, {\hat f}\right]=i \widehat{(X_rf)},
\end{eqnarray}

\noindent combined with (3.10). The space ${\cal H}$ is indeed
irreducible with respect to the family of operators $\{{\hat f},
V(\cdot)\}$ or equivalently $\{{\hat f},{\hat J}_r\}$.

We can express `functions of position' also via unitary operators
in the Weyl spirit as follows: for each real $f\in{\cal F} (G)$,
we define the unitary operator $U(f)$ by
     \begin{eqnarray}
     U(f) = e^{i\hat{f}} : (U(f)\psi)(g) =
     e^{if(g)}\psi(g).
     \end{eqnarray}
     \noindent
     It is then easy to see that we have the relations
     \begin{eqnarray}
     (U(f) V(g^{\prime})\psi)(g) &=&
     e^{if(g)}\psi\left(g^{\prime -1}g\right),\nonumber\\
     (V(g^{\prime}) U(f) \psi)(g) &=&
     e^{if(g^{\prime -1}g)}\psi\left(g^{\prime
     -1}g\right),\nonumber\\
     \left(U(f) V(g^{\prime}) (V(g^{\prime})U(f))^{\dag}
     \psi\right)(g)&&\nonumber\\
     &=&e^{if(g)-if(g^{\prime -1}g)}\psi(g) ,
     \end{eqnarray}

     \noindent
     which is in the spirit of eqns.(2.2,18), except that $f$ is
     not restricted to be linear in any coordinate variables.

 We see here that unlike in the $n$-dimensional Cartesian
case the `canonical momenta' are a non commutative system.
Therefore the analogue or generalisation of the single momentum
eigenstate $|{\underline p}>$ in the Cartesian situation will turn
out to be a generally multi-dimensional hermitian irreducible
representation of (3.10), namely the generators of some unitary
irreducible representation (UIR) of $G$. We will see this in
detail as we proceed.

For completeness we should mention the operators giving the right
regular representation of $G$. These are, say, ${\tilde V}(g)$,
defined by and obeying
\begin{eqnarray}
({\tilde V}(g^\prime )\psi )(g) &=& \psi(gg^\prime),\nonumber\\
{\tilde V}(g^\prime ) ~|g>&=&|gg^{\prime -1}>;\nonumber\\
{\tilde V}(g^\prime )~{\tilde V}(g)&=&{\tilde V}(g^{\prime}g );\nonumber\\
V(g^\prime )~{\tilde V}(g)&=&{\tilde V}(g)~V(g^\prime ).
\end{eqnarray}

\noindent However as is well known their generators ${\hat {\tilde
J}}_r$ are determined by ${\hat J}_r$  and the matrices (${\cal
D}_{r}^{s}(g)$) of the adjoint representation of $G$, by
\begin{eqnarray}
{\hat {\tilde J}}_r = -{\cal D}_{r}^{s}(g)~{\hat J}_s.
\end{eqnarray}

\noindent
Therefore it suffices to regard the collection of
operators $\{{\hat f},V(\cdot)\}$ as providing the replacement for
the Heisenberg-Weyl system in the present case.

Complementary to the `position' basis $|g>$ for ${\cal H}$ is a
`momentum' basis. This can be set up using the Peter-Weyl theorem
involving all the UIR's of $G$. We denote the various UIR's by
$j$, with dimension $N_j$; we label rows and columns within the
$j^{\mbox{th}}$ UIR by magnetic quantum numbers $m,n$. Thus the
unitary matrix representing $g\in G$ in the $j^{\mbox{th}}$ UIR is
\begin{eqnarray}
g \rightarrow (D_{mn}^{j}(g)).
\end{eqnarray}

In general each of $j,m,n$ is a collection of several independent
discrete quantum numbers, and there is a freedom of unitary
changes in the choice of $m,n$. In addition to unitarity and the
composition law :
\begin{eqnarray}
\sum\limits_{n} D^j_{mn}(g)^{*} D^j_{m^{\prime}n} (g) &=&
\delta_{mm^\prime},\nonumber\\
\sum\limits_{n} D^j_{mn}(g^{\prime}) D^j_{nn^\prime} (g) &=&
D^j_{mn^\prime}(g^{\prime}g),
\end{eqnarray}

\noindent we have orthogonality and completeness properties:
\begin{eqnarray}
\int\limits_G dg\; D^j_{mn}(g) D^{j^\prime}_{m^{\prime}n^{\prime}}
(g)^* &=& \delta_{jj^\prime}
\delta_{mm^\prime}\delta_{nn^\prime} /N_j~,\nonumber\\
\sum\limits_{jmn} N_j D^j_{mn} (g) D^j_{mn}(g^{\prime})^*
&=&\delta\left(g^{-1} g^{\prime}\right).
\end{eqnarray}

\noindent Then a simultaneous complete reduction of both
representations $V(\cdot), {\tilde V}(\cdot)$ of $G$ is achieved
by passing to a new orthonormal basis $|jmn>$ for ${\cal H}$.
Its definition and basic properties are :
\begin{eqnarray}
|jmn\rangle &=& N^{1/2}_j \int\limits_G
dg D^j_{mn}(g)|g\rangle , \nonumber\\
 \langle j^{\prime}
m^{\prime}n^{\prime}|jmn\rangle &=& \delta_{j^{\prime}j}
\delta_{m^{\prime}m} \delta_{n^{\prime}n} ; \nonumber\\
V(g)~ |jmn \rangle &=&\sum\limits_{m^\prime}  D^j_{mm^\prime}
(g^{-1})
|jm^\prime n>,\nonumber\\
{\tilde V}(g)~ |jmn \rangle &=&\sum\limits_{n^\prime}
D^j_{n^\prime n} (g) |jm n^\prime>.
\end{eqnarray}

\noindent Therefore in $|jmn>$ the index $n$ counts the
multiplicity of occurrence of the $j^{\mbox{th}}$  UIR in the
reduction of $V(\cdot)$ and $m$ performs a similar function in the
reduction of ${\tilde V}(\cdot)$.

We now regard the sets of $N_j^2$ states $\{|jmn>\}$ for each
fixed $j$ as `momentum eigenstates' in the present context. This
means that the $n$-dimensional real momentum eigevalue
${\underline p}$ in Cartesian quantum mechanics is now replaced by
a collection of (discrete) quantum numbers $jmn$. A vector
$|\psi>\in {\cal H}$ with wave function $\psi(g)$ is given in the
momentum description by a set of expansion coefficients
$\psi_{jmn}$:
\begin{eqnarray}
\psi \in~{\cal H}~\rightarrow \psi_{jmn} &=&
\langle jmn|\psi\rangle \nonumber\\
 &=& N_j^{1/2} \int\limits_G\; dg \;D^{j}_{mn} (g)^* \psi(g) ,\nonumber\\
\parallel\psi\parallel^2 &=& \sum\limits_{jmn} |\psi_{jmn}|^2 .
\end{eqnarray}

\noindent A normalised $|\psi>$ then determines two complementary
probability distributions, $|\psi(g)|^2$ on $G$ and
$|\psi_{jmn}|^2$ on `momentum space'.

In this situation a (provisional and overcomplete )  Wigner
distribution ${\tilde W}(g;~jmn~ m^\prime n^\prime)$ can be
defined for each $|\psi> \in {\cal H}$ ( or for any mixed state
${\hat \rho}$ as well). ( Here we depart slightly from the
notation in \cite{21}, so that our later expressions are more
concise). It transforms in a reasonable manner when $|\psi>$ is
acted upon by $V(\cdot)$ or ${\tilde V}(\cdot)$; and it reproduces
in a simple and direct way the two probability distributions
determined by $|\psi>$, as marginals. We give only the latter
property here:
\begin{eqnarray}
\sum\limits_{jmn} {\tilde W}(g; jmn\;mn) &=& |\psi(g)|^2 ,\nonumber\\
\int\limits_{G} dg {\tilde W}(g; jmn\;m^{\prime}n^{\prime}) &=&
\psi_{jm^\prime n^\prime}\;\psi_{jmn}^{*}.
\end{eqnarray}

\noindent The right hand side of the second relation is a natural
generalisation of $|\psi_{jmn}|^2$, to allow for freedom in the
choice of labels $m,n$ within each UIR $j$. The expression for
this Wigner distribution involves a function $s:~G\times
G\rightarrow G$ obeying certain conditions and is
\begin{eqnarray}
{\tilde W}(g; jmn\;m^{\prime}n^{\prime}) &=& N_j\int\limits_G
dg^{\prime} \int\limits_G dg^{\prime\prime}\;
\psi(g^{\prime\prime})\psi(g^{\prime})^*\;
D^j_{m^{\prime}n^{\prime}}(g^{\prime\prime})^* D^j_{mn}
(g^{\prime}) \;\delta\left( g^{-1}
s(g^{\prime},g^{\prime\prime})\right).
\end{eqnarray}

\noindent
Reality in the Cartesian or single angle-angular momentum
cases is replaced here by hermiticity :
\begin{eqnarray}
{\tilde W}(g;jmn\;m^{\prime}n^{\prime})^* &=&{\tilde W}
(g;jm^{\prime}n^{\prime}\;mn).
\end{eqnarray}

\noindent The conditions on $s(g^{\prime},g^{\prime\prime})$ to
ensure that all the above properties are secured are:
\begin{eqnarray}
g^{\prime},g^{\prime\prime}\;\in \;G \rightarrow
s(g^{\prime},g^{\prime\prime})&=& s(g^{\prime\prime},g^{\prime})\;
\in \; G ;\nonumber\\
s\left(g_1 g^{\prime} g_2 , g_1 g^{\prime\prime} g_2\right) &=&
g_1 \;s(g^{\prime},
g^{\prime\prime}) g_2 ,\nonumber\\
s(g^{\prime}, g^{\prime}) &=& g^{\prime}.
\end{eqnarray}

\noindent
We can simplify the problem of constructing such a
function by exploiting the second of these relations to write
\begin{eqnarray}
s(g^{\prime},g^{\prime\prime})=g^{\prime}s(e,g^{\prime
-1}g^{\prime\prime}) =g^{\prime}s_0(g^{\prime -1}g^{\prime\prime}),
\end{eqnarray}

\noindent
 so the function $s_0(g)$ of a single group element has
to satisfy
\begin{eqnarray}
&& s_0(e) = e ,\nonumber\\
&&s_0(g^{-1}) = g^{-1} s_0(g)=s_0(g)g^{-1}, \nonumber\\
&&s_0(g^{\prime} g g^{\prime -1}) = g^{\prime}\;s_0(g) g^{\prime
-1} .
\end{eqnarray}

\noindent The solution proposed in \cite{21} is to take $s_0(g)$ to
be the mid point along the geodesic from the identity $e\in G$ to
$g$. These geodesics are determined starting from the invariant
Cartan-Killing metric on $G$, and have the necessary behaviours
under left and right group actions to ensure that all of eqs.
(3.26,3.28) are obeyed. In the exponential notation of eq.(3.8) we
have
\begin{eqnarray}
s_0(\exp(\alpha^r~e_r))=\exp(\frac{1}{2}\alpha^r~e_r),
\end{eqnarray}

\noindent
 since it is true that geodesics passing through the
identity are one parameter subgroups. With this explicit
construction we have the additional relation
\begin{eqnarray}
s_0(g^{-1})&=& s_0(g)^{-1}, \nonumber\\
\mbox{ie.,}\;\;s_0(g)~s_0(g)&=&g.
\end{eqnarray}

\noindent
 Thus $s_0(g)$ is the (almost everywhere unique) square
root of $g$ and $s(g^\prime,g^{\prime\prime})$ is a kind of
symmetric square root of $g^\prime$ and $g^{\prime\prime}$.

We shall explore the properties of ${\tilde W}(g;~jmn~m^\prime
n^\prime)$ in the next Section, especially the sense in which it
contains information about $|\psi><\psi|$ in an overcomplete
manner. This will then lead to the Wigner-Weyl isomorphism for
quantum mechanics on a (compact simple) Lie group.

\section{ The Wigner-Weyl isomorphism in the Lie group case}

The definition (3.24) can be immediately extended to associate an
object ${\tilde W}_{{\hat A}}(g;~jmn~m^\prime n^\prime)$ with
every linear operator ${\hat A}$ on ${\cal H}$ ( of
Hilbert-Schmidt class). In terms of the integral kernel
$<g^{\prime\prime}|{\hat A}|g^\prime>$ of ${\hat A}$ we have
  \begin{eqnarray}
{\tilde W}_{\hat A}(g; jmn\;m^{\prime}n^{\prime}) &=&
N_j\int\limits_G dg^{\prime} \int\limits_G dg^{\prime\prime}\;
<g^{\prime\prime}|{\hat A}|g^{\prime}>
D^j_{m^{\prime}n^{\prime}}(g^{\prime\prime})^* D^j_{mn}
(g^{\prime})~\delta\left(
g^{-1}s(g^{\prime},g^{\prime\prime})\right).
\end{eqnarray}

\noindent
 It is indeed the case that this expression describes or determines
${\hat A}$ completely, however this happens in an over complete
manner. There are certain linear relations obeyed by ${\tilde
W}_{{\hat A}}(g;~jmn~m^\prime n^\prime)$ which have an ${\hat A}$
independent form. We now obtain these relations, then proceed to
construct a simpler expression which contains complete information
about ${\hat A}$ without redundancy.

The Dirac delta function in the integral on the right in eq.(4.1)
means that the only contributions to the integral are from the
points where
  \begin{eqnarray}
s(g^{\prime},g^{\prime\prime})=g.
  \end{eqnarray}

\noindent
Writing this as

  \begin{eqnarray}
s_0(g^{\prime -1}g^{\prime\prime})=g^{\prime -1}g,
  \end{eqnarray}

\noindent and then using eq.(3.30), we see that, say, in the
$g^{\prime\prime}$ integration the delta function picks out the
single point determined by
  \begin{eqnarray}
g^{\prime -1}g^{\prime\prime}&=&(g^{\prime -1}g)^2\nonumber\\
\mbox{ie},~~~~g^{\prime\prime}&=&gg^{\prime -1}g.
  \end{eqnarray}

\noindent This means that
$\delta(g^{-1}s(g^\prime,g^{\prime\prime}))$ is some Jacobian
factor times $\delta(g^{\prime\prime-1}gg^{\prime-1}g)$ . We are
therefore permitted to use this `value' for $g^{\prime\prime}$
elsewhere in the integrand, so
  \begin{eqnarray}
{\tilde W}_{\hat A}(g; jmn\;m^{\prime}n^{\prime}) =
N_j\int\limits_G dg^{\prime} \int\limits_G dg^{\prime\prime}\;
<gg^{\prime -1}g|{\hat A}|g^{\prime}>&&
D^j_{m^{\prime}n^{\prime}}(gg^{\prime -1}g)^* D^j_{mn}
(g^{\prime})\times \nonumber\\&& \delta\left(
g^{-1}s(g^{\prime},g^{\prime\prime})\right).
  \end{eqnarray}

\noindent
Transferring the $g$-dependent representation matrices
from the right to the left and using unitarity, we get:
  \begin{eqnarray}
&&\sum_{m^\prime n^\prime} D^j_{m^{\prime}m^{\prime\prime}}(g)
 D^j_{n^{\prime\prime}n^\prime}(g)
{\tilde W}_{\hat A}(g; jmn\;m^{\prime}n^{\prime})\nonumber\\
&=& N_j\int\limits_G dg^{\prime} \int\limits_G dg^{\prime\prime}\;
\delta\left( g^{-1}s(g^{\prime},g^{\prime\prime})\right)
<gg^{\prime -1}g|{\hat A}|g^{\prime}> D^j_{n^{\prime
\prime}m^{\prime \prime}}(g^{\prime}) D^j_{mn} (g^{\prime}).
  \end{eqnarray}

\noindent It is now clear we have symmetry of the expression on
the left hand side under the simultaneous interchanges
$m\leftrightarrow n^{\prime\prime}$ ,  $ n\leftrightarrow
m^{\prime\prime}$, a statement independent of ${\hat A}$. This is
the sense in which ${\tilde W}_{{\hat A}}(g;~jmn~m^\prime
n^\prime)$ contains information about ${\hat A}$ in an
overcomplete manner, and this happens only when $G$ is nonabelian.

Taking advantage of this, we now associate to ${\hat A}$ the
simpler quantity
  \begin{eqnarray}
W_{\hat A}(g; jmm^{\prime})&=&N_{j}^{-1}\sum_{n}
{\tilde W}_{\hat A}(g; jmn\;m^{\prime}n)\nonumber\\
&=&\int\limits_G dg^{\prime} \int\limits_G dg^{\prime\prime}\;
<g^{\prime\prime}|{\hat A}|g^{\prime}> D^j_{m
m^{\prime}}(g^{\prime}g^{\prime\prime -1}) \delta\left(
g^{-1}s(g^{\prime},g^{\prime\prime})\right).
  \end{eqnarray}

\noindent
We shall call this the Weyl symbol corresponding to the
operator ${\hat A}$. The passage ${\hat A}\rightarrow {\hat
A}^\dagger$ results in
  \begin{eqnarray}
W_{{\hat A}^\dagger}(g; jmm^{\prime}) =W_{\hat A}(g;
jm^{\prime}m)^*~.
  \end{eqnarray}

\noindent
It is easy to obtain the transformation properties of the
Weyl symbol under conjugation of ${\hat A}$ by either the left or
the right regular representation:
  \begin{eqnarray}
{\hat A}^\prime &=& V(g_0){\hat A}V(g_0)^{-1}:\nonumber\\
&& W_{{\hat A}^\prime}(g; jmm^{\prime})= \sum_{m_{1}m_{1}^\prime}
D^j_{m m_1}(g_0) D^j_{m^\prime m_{1}^{\prime}}(g_0)^*
~W_{{\hat A}}(g_{0}^{-1}g; jm_1m_{1}^{\prime}); \nonumber\\
{\hat A}^{\prime\prime} &=& {\tilde V}(g_0){\hat A}
{\tilde V}(g_0)^{-1}:\nonumber\\
&& W_{{\hat A}^{\prime\prime}}(g; jmm^{\prime})= W_{{\hat
A}}(gg_0; jmm^{\prime}).
  \end{eqnarray}

\noindent Next we can verify that if ${\hat A}$ and ${\hat B}$ are
any two Hilbert-Schmidt operators on ${\cal H}$, then
$\mbox{Tr}({\hat A}{\hat B})$ can be simply expressed in terms of
their Weyl symbols:
  \begin{eqnarray}
\mbox{Tr}({\hat A}~{\hat B})= \sum_{jmm^{\prime}}
N_{j}\int\limits_G dg~
 W_{\hat A}(g; jmm^\prime)~ W_{\hat B}(g; jm^\prime m).
  \end{eqnarray}

\noindent
 The proof exploits the completeness relation in (3.20) and the
 properties (3.26) of $s(g^\prime,g^{\prime\prime})$. This key result
 proves that ${\hat A}$ is indeed completely determined by its Weyl symbol :
 ${\hat A}$ is certainly determined by the values of
 $\mbox{Tr}({\hat A}{\hat B})$ for all ${\hat B}$, and the latter are known
 once the Weyl symbols are known.

 Before expressing the Weyl symbol of ${\hat A}$ in a form analogous to
 eq.(2.16), we give examples for some simple choices of ${\hat A}$:

\begin{eqnarray}
\begin{tabular}{ccc}
 $\underline{\hat{A}} $&\hspace{2cm} & $
\underline{W_{\hat{A}}(g; jmm^\prime)}$
\\
$\hat{f}=\int_{G} dg f(g)~|g><g|$ &\hspace{2cm} &
$f(g)\delta_{mm^\prime}$
\\
$V(g_0)$&\hspace{2cm}&$D^j_{m m^{\prime}}(g_{0}^{-1})$\\
${\tilde V}(g_0)$&\hspace{2cm}&$D^j_{m m^{\prime}}(gg_{0}g^{-1})$\\
${\hat f}~ V(g_0)$&\hspace{2cm}&$f(s_0(g_0)g)
D^j_{mm^{\prime}}(g_{0}^{-1})$\\
$ V(g_0)~{\hat f}$&\hspace{2cm}&$f(s_0(g_0)^{-1}g)
D^j_{mm^{\prime}}(g_{0}^{-1})$
\end{tabular}
\end{eqnarray}

We shall comment later on the structure of these Weyl symbols.
However it is already instructive to compare these results with
the Cartesian situation
\begin{eqnarray}
\begin{tabular}{ccc}
 $\underline{\hat{A}} $&\hspace{2cm} & $\underline{W(q,p)}$ \\
$\hat{f}=f({\hat q})$ &\hspace{2cm} & $f(q)$\\
$V(q^\prime)$&\hspace{2cm}&$\exp(-ipq^\prime)$\\
${\hat f}~ V(q^\prime)=f({\hat q})V(q^\prime)$&\hspace{2cm}&
$f(q+q^\prime/2)exp(-ipq^\prime)$\\
$ V(q^\prime)~f({\hat
q})$&\hspace{2cm}&$f(q-q^\prime/2)exp(-ipq^\prime) $
\end{tabular}
\end{eqnarray}

Now we turn to the problem of expressing the Weyl symbol of ${\hat
A}$ in the form
  \begin{eqnarray}
W_{\hat{A}}(g; jmm^\prime)= \mbox{Tr}({\hat A}~ {\hat W}(g;
jmm^\prime))
  \end{eqnarray}

\noindent for a suitable operator ${\hat W}(g;~jmm^\prime)$.  This
would be the analogue of ${\hat W}(q,p)$ in eq. (2.10). Since the
kernel $<g^{\prime\prime}|{\hat A}|g^\prime>$ is quite general, 
eq.(4.13) and eq.(4.7) imply :
  \begin{eqnarray}
<g^\prime|{\hat W}(g;~jmm^\prime)|g^{\prime\prime}>&=&
D^j_{mm^{\prime}}(g^\prime g^{\prime\prime-1})
\delta\left( g^{-1}s(g^{\prime},g^{\prime\prime})\right)\nonumber\\
&=&D^j_{mm^{\prime}}(g^\prime g^{\prime\prime-1}) \delta\left(
g^{-1}s_0(g^{\prime\prime}g^{\prime -1})g^{\prime}\right).
  \end{eqnarray}

\noindent
 We shall synthesise ${\hat W}(g;~jmm^\prime)$ in steps. We begin
 by defining a family of commuting operators $U(jmn)$ in the manner of
 eq.(3.5), all of them diagonal in the `position' basis :
\begin{eqnarray}
 (U(jmn)\psi)(g)=D^j_{mn}(g)\psi(g).
 \end{eqnarray}

 \noindent
  These are analogous to the Cartesian $U(p^\prime)$:
  labelled by a `momentum
eigenvalue' $jmn$, functions of `position' alone. They are unitary
in the matrix sense:
\begin{eqnarray}
\sum_{m}~U(jmn)^\dagger~U(jmn^\prime)=\sum_{m}
U(jnm)^\dagger~U(jn^\prime m) =\delta_{n^\prime n}~ I.
 \end{eqnarray}

These operators allow us to express the map $f\in {\cal F}(G)
\rightarrow \hat{f}$ of eq.(3.5) more explicitly as follows:
\begin{eqnarray}
f(g) = \sum\limits_{jmn}\;f_{jmn}\;D^j_{mn}(g) \Rightarrow \hat{f}
= \sum\limits_{jmn}\;f_{jmn}\;U(jmn) .
\end{eqnarray}

\noindent Upon conjugation by $V(g)$ we have
     \begin{eqnarray}
     V(g)^{-1} U(jmn) V(g) = \sum\limits_{m^{\prime}}\;
     D^j_{mm^{\prime}}(g)\;U(jm^{\prime}n).
     \end{eqnarray}

\noindent Combining eqs.(3.16, 3.20, 4.15) we easily obtain the
trace orthonormality property

     \begin{eqnarray}
     \mbox{Tr}\left((U(j^{\prime}m^{\prime}n^{\prime})
     V(g^{\prime}))^{\dag}\;U(jmn)\;V(g)\right) =
     N^{-1}_j \delta_{j^{\prime}j} \delta_{m^{\prime}m}
     \delta_{n^{\prime}n} \delta\left(g^{-1}g^{\prime}\right) ,
     \end{eqnarray}

     \noindent
     analogous to eqs.(2.7,2.5).  The action of
     $U(j^{\prime}m^{\prime}n^{\prime})$ on the `momentum
     eigenstates' $|jmn>$ can be worked out; it involves the
     Clebsch-Gordan coefficients for the reduction of direct
     products of two general UIR's of $G$ and reads:
          \begin{eqnarray}
          U(j^{\prime}m^{\prime}n^{\prime}) |jmn> =
          \sum_{j^{\prime\prime}m^{\prime\prime}
          n^{\prime\prime}\lambda}\sqrt{\frac{N_{j}}
          {N_{j^{\prime\prime}}}}\;
          C_{m^{\prime}m m^{\prime\prime}}^{j^{\prime}j
          j^{\prime\prime}\lambda}\;^*\;C_{n^{\prime}n
          n^{\prime\prime}}^{j^{\prime} j
          j^{\prime\prime}\lambda}\;
          |j^{\prime\prime}m^{\prime\prime}n^{\prime\prime}\rangle.
          \end{eqnarray}

          \noindent
          Here $\lambda$ is a multiplicity index keeping track of
          the possibly multiple occurrences of the UIR
          $D^{j^{\prime\prime}}$ in the reduction of the direct
          product $D^{j^{\prime}}\times D^{j}$.  The significance
          of this relation is similar in spirit to the statement
          in the Cartesian case that $U(p^{\prime}) =\exp\left(i
          p^{\prime}\hat{q}\right)$ generates a translation in
          $\hat{p}$, in other words that in the momentum
          description $\hat{q}$ is given by the differential
          operator $i\;\frac{d}{dp}$.  The result (4.20) however
          involves discrete labels since $G$ is compact, unlike
          continuous Cartesian variables, and incorporates
          nonabelianness as well.  Therefore `translating' the
          momentum $jmn$ by the `amount'
          $j^{\prime}m^{\prime}n^{\prime}$ yields several final
          momenta
          $j^{\prime\prime}m^{\prime\prime}n^{\prime\prime}$
          according to the contents of the direct product
          $D^{j^{\prime}}\times D^{j}$ of UIR's of $G$.

  Now multiply both sides of eq.(4.14) by $
D_{m_1m_{1}^\prime}^ {j_1}(g)$ and integrate with respect to $g$,
this is `Fourier transformation' with respect to $g$ and gives
  \begin{eqnarray}
<g^\prime|\int\limits_{G} dg~ D^{j_1}_{m_1m_{1}^{\prime}}(g) {\hat
W}(g;~jmm^\prime)|g^{\prime\prime}>= D^j_{mm^{\prime}}(g^\prime
g^{\prime\prime-1})~
D^{j_1}_{m_1m_{1}^{\prime}}(s_0(g^{\prime\prime}g^{\prime
-1})g^{\prime}).
  \end{eqnarray}

\noindent
Now perform an `inverse Fourier transformation' with
respect to the momenta $jmm^\prime$ to get
  \begin{eqnarray}
\sum_{jmm^\prime} \; N_j\;D^j_{mm^{\prime}}(g_1)^{*} &&
<g^\prime|\int\limits_{G} dg~ D^{j_1}_{m_1m_{1}^{\prime}}(g)~
{\hat W}(g;~jmm^\prime)|g^{\prime\prime}>,
\nonumber\\
&=&D^{j_1}_{m_1m_{1}^{\prime}}(s_0(g^{\prime\prime}g^{\prime
-1})g^{\prime})
~\delta(g_1 g^{\prime\prime}g^{\prime -1}),\nonumber\\
&=& D^{j_1}_{m_1m_{1}^{\prime}}(s_0(g_{1}^{-1})g^{\prime})
~\delta(g_1 g^{\prime\prime}g^{\prime -1}),\nonumber\\
&=&<g^\prime|g_1g^{\prime\prime}>~
D^{j_1}_{m_1m_{1}^{\prime}}(s_0(g_{1}^{-1})g^{\prime}),\nonumber\\
&=&\sum_{m_2}<g^\prime|U(j_1m_2m_{1}^\prime)~V(g_1)|g^{\prime\prime}>
~D^{j_1}_{m_1m_2}(s_0(g_{1}^{-1})).
\end{eqnarray}

\noindent Comparing the two sides and `peeling off' $<g^\prime|$
and $|g^{\prime \prime}>$ gives:
  \begin{eqnarray}
\sum_{jmm^\prime}\;N_j\; D^j_{mm^{\prime}}(g_1)^{*} &&
\int\limits_{G}dg~ D^{j_1}_{m_1m_{1}^{\prime}}(g)~ {\hat
W}(g;~jmm^\prime)=
\sum_{m_2}D^{j_1}_{m_1m_2}(s_0(g_{1}^{-1}))~U(j_1m_2m_{1}^\prime)~V(g_1).
  \end{eqnarray}

\noindent
Then `Fourier inversion' twice yields the result:
  \begin{eqnarray}
{\hat
W}(g;~jmm^\prime)=\sum_{j_1m_1m_2}~N_{j_1}\int\limits_{G}dg_1
U(j_1m_2m_{1})V(g_1)D^j_{mm^{\prime}}(g_1) D^{j_1}_{m_1m_2}
(g^{-1}s_0(g_{1}^{-1})).
  \end{eqnarray}

\noindent This may be compared in every detail with the Cartesian
result in eq.(2.10): the correspondence of arguments and
integration/summation variables is (including the factors
representing `momentum eigenfunctions')
  \begin{eqnarray}
&&q\rightarrow ~g,~p\rightarrow~jmm^{\prime},~ q^\prime
\rightarrow g_1,
~p^\prime \rightarrow~j_1m_1m_2;\nonumber\\
&& e^{ipq^\prime}\rightarrow ~D^j_{mm^{\prime}}(g_1),~~
e^{-ip^\prime(q+q^\prime/2)}\rightarrow ~D^{j_1}_{m_1m_2}
(g^{-1}s_0(g_{1}^{-1})).
  \end{eqnarray}

\noindent
Giving due attention to the new matrix features, the
correspondence is quite remarkable.

Combining eqs.(3.28,4.14) we obtain the relation
 \begin{eqnarray}
 \hat{W}(g; jmn)^{\dag} = \hat{W}(g; jnm) .
 \end{eqnarray}

 \noindent
 Similarly combining eqs.(4.24, 4.19) and carrying out quite
 elementary operations leads to analogues to the Cartesian
 relations (2.11, 2.15) in the forms

      \[\mbox{Tr} \left(\hat{W}(g^{\prime};
      j^{\prime}m^{\prime}n^{\prime})^{\dag}\;\hat{W}(g;jmn)\right)=
      N_j^{-1}
      \delta_{jj^{\prime}}\delta_{mm^{\prime}}
      \delta_{nn^{\prime}}\delta\left(g^{-1}\;g^{\prime}\right)
      ,\]
  \begin{eqnarray}
  \hat{A} = \sum\limits_{jmn}\;N_j\;\int\limits_{G}\;dg\;W_{\hat{A}}
  (g;jnm)\;\hat{W}(g;jmn).
  \end{eqnarray}

\noindent
We may thus conclude that we have succeeded in setting up
a Wigner-Weyl isomorphism for quantum mechanics on a compact
simple Lie group with reasonable properties.

\section{The star product for Weyl symbols}

In this Section we sketch the derivation of the expression for
noncommutative operator multiplication in terms of the
corresponding Weyl symbols, relegating some details to the
Appendix.  Thus, for two operators $\hat{A}$ and $\hat{B}$ we seek
an expression for the Weyl symbol of $\hat{A}\hat{B}$ in terms of
those of $\hat{A}$ and $\hat{B}$ in the form
\begin{eqnarray}
W_{\hat{A}\hat{B}} (g;j m n) = \left(W_{\hat{A}} *
W_{\hat{B}}\right) (g; j m n).
\end{eqnarray}

\noindent From eq.(4.13) we have
\begin{eqnarray}
\left(W_{\hat{A}} * W_{\hat{B}}\right) (g; j m n) =
\mbox{Tr}\left(\hat{A}\hat{B}\hat{W} (g; j m n)\right),
\end{eqnarray}

\noindent so using eq.(4.27) for $\hat{A}$ as well as for
$\hat{B}$ we have
\begin{eqnarray}
\left(W_{\hat{A}} * W_{\hat{B}}\right) (g; j m n) &=&
\sum_{\stackrel{j^{\prime}m^{\prime}n^{\prime}}
{j^{\prime\prime}m^{\prime\prime}n^{\prime\prime}}}\;N_{j^\prime}\;
N_{j^{\prime\prime}}
\int\limits_{G}\;dg^{\prime\prime}\;\int\limits_{G}\;
dg^{\prime}\;W_{\hat{A}}\left(g^{\prime\prime};j^{\prime\prime}\;
n^{\prime\prime}\;m^{\prime\prime}\right) W_{\hat{B}}
\left(g^{\prime};j^{\prime} \;n^{\prime}\;m^{\prime}\right)
\times\nonumber\\
&&\mbox{Tr}\left(\hat{W}\left(g^{\prime\prime};j^{\prime\prime}\;
m^{\prime\prime}\;n^{\prime\prime}\right)\; \hat{W}
\left(g^{\prime};j^{\prime}\;m^{\prime}\;n^{\prime}\right)\;
\hat{W}(g; j m n)\right).
\end{eqnarray}

\noindent We therefore need to compute the trace of the product of
three $\hat{W}$'s, which is a nonlocal integral kernel defining
the (associative but noncommutative) star product on the left hand
side.  The two ingredients for this calculation are expressions
for the product $U(jmn)\;V(g)$ in terms of
$\hat{W}(g^{\prime};j^{\prime}m^{\prime}n^{\prime})$, and for the
product $U(j^{\prime}m^{\prime}n^{\prime})\;V(g^{\prime})\;
U(jmn)\;V(g)$ in terms of similar products $UV$.  These are:
\begin{mathletters}
\begin{eqnarray}
U(jmn)\;V(g) = \sum\limits_{j^{\prime}m^{\prime}n^{\prime}}\;
N_{j^{\prime}}\;D^{j^{\prime}}_{m^{\prime}n^{\prime}}(g)^*
\int\limits_{G}\;dg^{\prime}\;D^j_{mn}(s_0(g)g^{\prime})\;
\hat{W}(g^{\prime};j^{\prime}m^{\prime}n^{\prime}),\\
U(j^{\prime}m^{\prime}n^{\prime})\;V(g^{\prime})\;U(jmn)\;V(g) =
\sum\limits_{j^{\prime\prime}m^{\prime\prime}n^{\prime\prime}k}\;
C^{j^{\prime}}_{m^{\prime}n^{\prime}}\;^{j}_{kn}\;^{j^{\prime\prime}}
_{m^{\prime\prime}n^{\prime\prime}}\;D^{j}_{mk}\;\left(g^{\prime
-1}\right) \times\nonumber\\
U(j^{\prime\prime}m^{\prime\prime}n^{\prime\prime})\;V(g^{\prime}g).
\end{eqnarray}
\end{mathletters}

\noindent The derivations are given in the Appendix, and the
$C$-symbol on the right in the second equation is a sum of products
of Clebsch-Gordan coefficients of the type occurring in eq.(4.20).

Starting from eq.(4.24) and using eq.(5.4b) we have for the
product of two $\hat{W}$'s:
\begin{eqnarray*}
\hat{W}(g^{\prime};j^{\prime}m^{\prime}n^{\prime})\;\hat{W}(g;jmn)
=\sum\limits_{\stackrel{j_0,m_0,n_0}{j_0^{\prime}m^{\prime}_0n^{\prime}_0}}\;
N_{j_{0}}\;N_{j^{\prime}_{0}}\;\int\limits_{G}\;dg_0\;\int\limits_{G}\;
dg^{\prime}_0\;D^j_{mn}
(g_0)\;D^{j^{\prime}}_{m^{\prime}n^{\prime}}
\left(g^{\prime}_0\right) \times
\end{eqnarray*}
\begin{eqnarray*}
D^{j_{0}}_{n_{0}m_{0}}\left(g^{-1}
s_0\left(g^{-1}_{0}\right)\right)\;
D^{j^{\prime}_{0}}_{n_{0}^{\prime}m^{\prime}_0} \left(g^{\prime
-1} s_0\left(g^{\prime -1}_{0}\right)\right)\;
U\left(j^{\prime}_{0}m^{\prime}_{0}n^{\prime}_{0}\right)\;
V\left(g^{\prime}_{0}\right)\; U(j_0m_0n_0)\;V(g_0)
\end{eqnarray*}
\begin{eqnarray*}
=\sum\limits_{\buildrel {\stackrel{j_{0}m_{0}n_{0}k_{0}}
{j^{\prime}_{0}m^{\prime}_{0}n^{\prime}_{0}}}
\over{j^{\prime\prime}_{0}m^{\prime\prime}_{0}n^{\prime\prime}_{0}}}
\;N_{j_{0}}\;N_{j^{\prime}_{0}}\;
C^{j^{\prime}_{0}}_{m^{\prime}_{0}n^{\prime}_{0}}\;^{j_{0}}_{k_{0}n_{0}}
\;^{j^{\prime\prime}_{0}}_{m^{\prime\prime}_{0}n^{\prime\prime}_{0}}\;
\int\limits_{G}\; dg_0\;\int\limits_{G}\;dg^{\prime}_{0}\;
D^{j}_{mn}(g_0)\;D^{j^{\prime}}_{m^{\prime}n^{\prime}}\left(g_{0}^{\prime}
\right) \times
\end{eqnarray*}
\begin{eqnarray*}
D^{j_{0}}_{m_{0}k_{0}}\left(g^{\prime -1}_{0}\right)\;
D^{j_{0}}_{n_{0}m_{0}}\left(g^{-1}
s_0\left(g_{0}^{-1}\right)\right)\;D^{j^{\prime}_{0}}_{n_{0}^{\prime}
m^{\prime}_{0}}\left(g^{\prime -1} s_0\left(g^{\prime
-1}_{0}\right)\right)\times
\end{eqnarray*}
\begin{eqnarray}
U\left(j^{\prime\prime}_{0} m^{\prime\prime}_{0}
n^{\prime\prime}_{0}\right)\; V\left(g^{\prime}_{0}\;g_0\right) .
\end{eqnarray}

\noindent If here we use eq.(5.4a) and then eq.(4.27) we obtain
for the kernel in eq.(5.3):
\begin{eqnarray*}
\mbox{Tr}\left(\hat{W}(g^{\prime\prime};
j^{\prime\prime}m^{\prime\prime}n^{\prime\prime})
\hat{W}(g^{\prime};
j^{\prime}m^{\prime}n^{\prime})\hat{W}(g;jmn)\right) =
\sum\limits_{\buildrel {\stackrel{j_{0}m_{0}n_{0}k_{0}}
{j^{\prime}_{0}m^{\prime}_{0}n^{\prime}_{0}}}
\over{j^{\prime\prime}_{0}m^{\prime\prime}_{0}n^{\prime\prime}_{0}}}
N_{j_{0}}\;N_{j^{\prime}_{0}}\;
C^{j^{\prime}_{0}}_{m^{\prime}_{0}n^{\prime}_{0}}\;^{j_{0}}_{k_{0}n_{0}}
\;^{j^{\prime\prime}_{0}}_{m^{\prime\prime}_{0}n^{\prime\prime}_{0}}\;
\times
\end{eqnarray*}

\begin{eqnarray*}
\int\limits_{G}\; dg_0\;\int\limits_{G}\;dg^{\prime}_{0}\;
D^{j}_{mn}(g_0)\;D^{j^{\prime}}_{m^{\prime}n^{\prime}}\left(g_{0}^{\prime}
\right)\;D^{j^{\prime\prime}}_{n^{\prime\prime}m^{\prime\prime}}
\left(g^{\prime}_0 g_{0}\right)^*\; D^{j_{0}}_{m_{0}k_{0}}
\left(g_{0}^{\prime -1}\right)\times
\end{eqnarray*}
\begin{eqnarray}
D^{j_{0}}_{n_{0}m_{0}}\left(g^{-1}\;s_0\left(g_0^{-1}\right)\right)\;
D^{j_{0}^{\prime}}_{n^{\prime}_{0}m^{\prime}_{0}} \left(g^{\prime
-1}s_{0}\left(g_{0}^{\prime -1}\right)\right)\;
D^{j_{0}^{\prime\prime}}_{m_{0}^{\prime\prime}
n_{0}^{\prime\prime}}\left(s_0\left(g_{0}^{\prime}g_{0}\right)
g^{\prime\prime}\right).
\end{eqnarray}

\noindent The star product of eq.(5.3) is then obtained by
inserting this integral kernel on the right hand side.

A slightly simpler expression - which amounts to trading four of
the $D$-functions for Dirac delta functions - results from direct
use of eq.(4.14):
\begin{eqnarray*}
\mbox{Tr}\left(\hat{W}(g^{\prime\prime};j^{\prime\prime}
m^{\prime\prime}n^{\prime\prime})
\hat{W}(g^{\prime};j^{\prime}m^{\prime}n^{\prime})
\hat{W}(g;jmn)\right)
=\int\limits_{G}\;dg_0\;\int\limits_{G}\;dg_{0}^{\prime}\;
\int\limits_{G}\;dg_0^{\prime\prime}\times
\end{eqnarray*}
\begin{eqnarray*}
\langle
g_0|\hat{W}(g^{\prime\prime};j^{\prime\prime}m^{\prime\prime}
n^{\prime\prime})|g^{\prime}_{0}\rangle \langle
g^{\prime}_{0}|\hat{W}(g^{\prime};j^{\prime}m^{\prime}n^{\prime})
|g^{\prime\prime}_{0}\rangle\langle g_0^{\prime\prime}|\hat{W}
(g;jmn)|g_0\rangle
\end{eqnarray*}
\begin{eqnarray*}
=\int\limits_{G}\;dg_0\;\int\limits_{G}\;dg^{\prime}_0\;
\int\limits_{G}\;dg^{\prime\prime}_{0}\;\;D^{j^{\prime\prime}}_{m^{\prime\prime}
n^{\prime\prime}}\left(g_0\;g_0^{\prime -1}\right)\;
D^{j^{\prime}}_{m^{\prime}n^{\prime}}\left(g^{\prime}_{0}\;g^{\prime\prime
-1}_{0}\right)\;D^j_{mn}\left(g^{\prime\prime}_{0}g^{-1}_{0}\right)\times
\end{eqnarray*}
\begin{eqnarray}
\delta\left(g^{\prime\prime
-1}s\left(g_0,g^{\prime}_{0}\right)\right)\; \delta\left(g^{\prime
-1}s\left(g_{0}^{\prime},g^{\prime\prime}_{0}\right)\right)\;
\delta\left(g^{-1}\;s\left(g^{\prime\prime}_{0},g_{0}\right)\right).
\end{eqnarray}

These expressions for the star product show an unavoidable
complexity for general compact nonabelian $G$.  In the one
dimensional abelian (but non Cartesian) case $Q=\mS^1$, there are
some simplifications.  Referring to Section II, we have the rule
for Weyl symbols given by eqns.(2.25,28):
\begin{eqnarray}
a(\theta;m) &=&
\mbox{Tr}\left(\hat{A}\;\hat{W}(\theta;m)\right),\nonumber\\
\hat{A}&=& 2\pi\;\sum\limits_{m\in \mZ}\;\int\limits^{\pi}_{-\pi}
d\theta\;a(\theta;m)\hat{W}(\theta;m).
\end{eqnarray}

\noindent The star product then appears as
\begin{eqnarray*}
(a \star b) (\theta;m)
=\sum\limits_{m^{\prime},m^{\prime\prime}\in
\mZ}\;\int\limits^{\pi}_{-\pi}
d\theta^{\prime\prime}\;\int\limits^{\pi}_{-\pi}
d\theta^{\prime}&&\mbox{Tr}\left(\hat{W}(\theta^{\prime\prime};
m^{\prime\prime})\hat{W}(\theta^{\prime};m^{\prime})
\hat{W}(\theta;m)\right)\times\\
&&a(\theta^{\prime\prime};m^{\prime\prime})\;
b(\theta^{\prime};m^{\prime}),
\end{eqnarray*}
\begin{eqnarray*}
\mbox{Tr}\left(\hat{W}(\theta^{\prime\prime};m^{\prime\prime})\;
\hat{W}(\theta^{\prime};m^{\prime})\;\hat{W}(\theta;m)\right)
=\frac{1}{4\pi^2}\;\sum\limits_{m_{0},m_{0}^{\prime}\in
\mZ}\;\int\limits^{\pi}_{-\pi}
d\theta^{\prime}_{0}\;\int\limits^{\pi}_{-\pi} d\theta_0\;
e^{\frac{i}{2}\left(m^{\prime}_{0}\theta_{0}-m_{0}
\theta^{\prime}_{0}\right)}\times
\end{eqnarray*}
\begin{eqnarray}
\exp\left[i\left(m\theta_{0}-m_{0}\theta +
m^{\prime}\theta^{\prime}_{0} -m^{\prime}_{0}\theta^{\prime}
+\left(m_0+m^{\prime}_0\right)
\theta^{\prime\prime}-m^{\prime\prime}
\left(\theta_0+\theta_0^{\prime}\right) \right)\right].
\end{eqnarray}

\noindent This expression for the kernel results from eq.(5.6) if
we first drop the magnetic quantum numbers
$m,n,m^{\prime},n^{\prime},m^{\prime\prime},n^{\prime\prime},m_0,n_0,
k_0,m^{\prime}_0, n^{\prime}_0,
m^{\prime\prime}_0,n^{\prime\prime}_0$; then set the
dimensionalities $N_{j_{0}}, N_{j_{0}^{\prime}}$ equal to unity;
next make the replacements $j\rightarrow m, j^{\prime}\rightarrow
m^{\prime}, j^{\prime\prime}\rightarrow m^{\prime\prime},
j_0\rightarrow m_0, j^{\prime}_0\rightarrow m^{\prime}_0,
g_0\rightarrow \theta_0, g^{\prime}_0\rightarrow
\theta^{\prime}_0$, and use for the $C$ coefficient the kronecker
delta $\delta_{j^{\prime\prime}_{0}, m_0 + m_0^{\prime}}$.  Even
with some simplifications, the kernel in eq.(5.9) remains nonlocal
because of (among other things) the occurrence of half angles in
the exponent.

\section{Discussion and concluding remarks}

The characteristic feature revealed by our analysis is that for
quantum mechanics on a Lie group $G$ as configuration space, the
concept of canonical momentum is a collection of non-commuting
operators ${\hat J}_r$, in fact constituting the Lie algebra of
the left regular representation of $G$ on $L^2(G)$. This in itself
is known, but it results in the analogues of `momentum eigenvalue'
being a set of discrete labels $jmn$, and the single Cartesian
momentum eigenvector $|{\underline p>}$ being replaced by a
multidimensional set of vectors $\{|jmn>\}$. Other consequences of
this non-abelianness should be noted. One needs to work with both
overcomplete and with complete non-redundant Weyl symbols for
general operators ${\hat A}$ : the former are useful for
reproducing in a simple manner the two complementary marginal
probability distributions associated with a pure or mixed quantum
state from its Wigner distribution as shown in eq.(3.23); while
the latter lead to the Wigner-Weyl isomorphism in a reasonable
manner.

It is interesting that the Weyl symbols $ W_{{\hat
A}}(g;~jmm^\prime)$ are not complex valued functions on the
classical phase space $T ^*G$. They may be more compactly viewed as
follows. Whereas by the Peter-Weyl theorem the Hilbert space
${\cal H}=L ^2(G)$ carries each UIR ${\cal D}^{(j )}(\cdot)$ of
$G$ as often as its dimension $N_j$, the structure of eq.(4.7)
leads us to define a `smaller' Hilbert space ${\cal H}_0$ carrying
each UIR of $G$ \underline{exactly once}:
\begin{eqnarray}
{\cal H}_0&=& \sum\limits_{j}\;_{\bigoplus}\;
{\cal H}^{(j)}\nonumber\\
{\cal H}^{(j)}& =& \mbox{Sp}~\{|jm)\},~~\mbox{dim}~{\cal
H}^{(j)}=N_j, \nonumber\\
(j^{\prime}m^{\prime}|jm)&=&
\delta_{j^{\prime}j}\;\delta_{m^{\prime}m},
\end{eqnarray}

\noindent with ${\cal H}^{(j)}$ carrying the UIR ${\cal D}^{(j
)}(\cdot)$ of $G$. Then the Weyl symbol of a general operator
${\hat A}$, $W_{{\hat A}}(g;~jmm^\prime) $, may be regarded as a
function of $g\in G$ and an operator on ${\cal H}_0$, but with the
crucial property that it is \underline{block~diagonal} with
respect to the decomposition (6.1) of ${\cal H}_0$. This is
evident from the examples of Weyl symbols given in eq.(4.11); in
the Cartesian case in eq.(4.12) such features are of course
absent. This can be understood also from the following point of
view. In the normal quantum description  an operator ${\hat A}$ on
${\cal H}=L^2(G)$ can be given via its kernel
$<g^{\prime\prime}|{\hat A}|g^\prime> $, or via its complementary
diagonal plus off diagonal matrix elements $<j^{\prime} m^{\prime}
n^\prime|{\hat A}|jmn> $. If in the latter we `trade' half the
labels for a dependence on a group element $g$, we arrive at the
Weyl symbol $W_{{\hat A}}(g;~jmm^\prime) $ viewed as a block
diagonal operator on ${\cal H}_0 $ with simultaneously a
dependence on $g$. Thus while the Wigner-Weyl isomorphism does not
work directly with the true classical phase space $T^*G$, it seems
to use what may be called a non-commutative cotangent space,
standing somewhere between $T^*G$ and operators on $L^2(G)$.

Nevertheless the link to functions on the classical phase space
$T^*G$ can be established, as we will see below.

We may use the phrase `semiquantised phase space' for the space on
which the Weyl symbols $W_{\hat{A}}(g;jmn)$ of operators $\hat{A}$
are defined.  It is to be understood that this phrase includes the
restriction that only ($g$-dependent) \underline{block-diagonal}
operators on ${\cal H}_0$ are encountered.  This may be viewed as
a superselection rule.  In detail, given an operator $\hat{A}$ on
${\cal H}=L^2(G)$, we associate with it the $g$-dependent
block-diagonal operator
\begin{eqnarray}
\tilde{A}(g) = \sum\limits_j\;\sum\limits_{m,n}\;\sqrt{N_j}~W_{\hat{A}}
(g;jmn)|jm)(jn|,
\end{eqnarray}

\noindent acting on ${\cal H}_0$, and we then have the connection
\begin{eqnarray}
\mbox{Tr}_{{\cal H}}(\hat{A}\hat{B}) = \int\limits_{G} dg\;
\mbox{Tr}_{{\cal H}_{0}} (\tilde{A}(g)\;\tilde{B}(g)).
\end{eqnarray}

\noindent The Weyl symbol $\tilde{A}(g)$ is simpler than $\hat{A}$
both in that it acts on the much smaller Hilbert space ${\cal
H}_{0}$, and in that it is block-diagonal.

To finally establish the link to suitable functions on the
classical phase space $T^*G$, we exploit both the fact that the
representation of $G$ on ${\cal H}_0$ has a multiplicity-free
reduction into UIR's, and the fact that $\tilde{A}(g)$ is
block-diagonal.  Let us denote the generators of $G$ on ${\cal
H}_0$ by $\hat{J}_{r}^{(0)}, \;r=1,2,\ldots,n$.  The Weyl symbol
$\tilde{A}(g)$ may initially be written as the direct sum of
symbols $\tilde{A}_j(g)$ acting within each subspace ${\cal
H}^{(j)}$ in ${\cal H}_0$:
\begin{eqnarray}
\tilde{A}(g) &=&
\sum\limits_{j}{_{\bigoplus}}\;\tilde{A}_j(g),\nonumber\\
\tilde{A}_j(g)&=& \sum\limits_{m,n}\;\sqrt{N_j}~W_{\hat{A}}(g;jmn)\;
|jm)(jn|.
\end{eqnarray}

\noindent Next, using the irreducibility of
$\{\hat{J}_{r}^{(0)}\}$ acting on ${\cal H}^{(j)}$, we can expand
$\tilde{A}_j(g)$  uniquely as a sum of symmetrised polynomials in
$\hat{J}_{r}^{(0)}$:
\begin{eqnarray*}
\tilde{A}_j(g) =\sum\limits_{N=0,1,\ldots}
\sum\limits_{r_{1},r_{2}\ldots r_{N}} a_{r_{1}\ldots r_{N}}(g;j)
\left\{\hat{J}_{r_{1}}^{(0)}\;\hat{J}_{r_{2}}^{(0)}\ldots
\hat{J}_{r_{N}}^{(0)}\right\}_S^{(j)},
\end{eqnarray*}
\begin{eqnarray}
\left\{\hat{J}^{(0)}_{r_{1}}\hat{J}^{(0)}_{r_{2}}\ldots
\hat{J}^{(0)}_{r_{N}}\right\}_S^{(j)}
=\frac{1}{N!}\;\sum\limits_{P\in
S_{N}}\left(\hat{J}^{(0)}_{r_{P(1)}}\ldots
\hat{J}^{(0)}_{r_{P(N)}}\right)^{(j)}.
\end{eqnarray}

\noindent Here the upper limit of $N$ is determined by the UIR
$D^j; S_N$ is the permutation group on $N$ symbols; and the
superscript $(j)$ denotes the restriction to ${\cal H}^{(j)}$. The
coefficients $a_{r_{1}\ldots r_{N}} (g;j)$ are $c$-number
quantities symmetric in $r_1\ldots r_N$.  If we now replace their
$j$ dependences by dependences on the independent mutually
commuting Casimir operators $\hat{{\cal C}}$ of $G$, themselves
symmetric homogeneous polynomials in $\hat{J}_{r}^{(0)}$, we can
use (6.5) in (6.4) and write:
\begin{eqnarray} \tilde{A}(g) =
\sum\limits^{\infty}_{N=0} \sum\limits_{r_{1}\ldots
r_{N}}\;a_{r_{1}\ldots r_{N}} \;(g;\hat{{\cal C}})
\left\{\hat{J}^{(0)}_{r_{1}}\ldots
\hat{J}^{(0)}_{r_{N}}\right\}_S.
\end{eqnarray}

\noindent This expression for the Weyl symbol $\tilde{A}(g)$ of
$\hat{A}$ can now be put into one-to-one correspondence with the
classical phase space function
\begin{eqnarray}
a(g;J) = \sum\limits^{\infty}_{N=0} \sum\limits_{r_{1}\ldots
r_{N}}\;a_{r_{1}\ldots r_{N}} (g;{\cal C})\;J_{r_{1}}\ldots
J_{r_{N}},
\end{eqnarray}

\noindent where the commuting classical variables $J_r$ are the
canonical momentum coordinates of the classical phase space $T^*G$
\cite{21}, while ${\cal C}$ are invariant (Casimir) homogeneous
polynomials in them.  Thus we have the two-stage sequence of
correspondences
\begin{eqnarray}
\hat{A}\;\mbox{on}\;{\cal H}=L^2(G)\Longleftrightarrow
\tilde{A}(g) = \mbox{block-diagonal operator on}\;{\cal H}_0
\longleftrightarrow a(g;J)\in {\cal F}(T^*G).
\end{eqnarray}

\noindent The importance of the multiplicity-free nature of the
representation of $G$ on ${\cal H}_0$, and the super selection
rule, is evident.  In contrast to the Cartesian case in Section
II, the appearance of the semi-quantised phase space as an
intermediate step is to be noted. We hope to return to this aspect 
in a future publication.

\newpage
\def\theequation{A.\arabic{equation}}
\appendix{\bf{Appendix }}
\setcounter{equation}{0}

We indicate here the derivations of eqs.(5,4a,b).  For eq.(5.4a),
we begin with eq.(4.23) and use the unitarity of the $D$-matrices
to shift the $D$-matrix on the right to the left.  This
immediately gives eq.(5.4a).  For eq.(5.4b) we begin with the
decomposition of the product of two $U$'s; from eq.(4.15), using
eq.(A.29) in \cite{21},
\begin{eqnarray*}
U(j^{\prime}m^{\prime}n^{\prime}) U(jmn) |g\rangle =
D^{j^{\prime}}_{m^{\prime}n^{\prime}}(g)\;D^j_{mn}(g) |g\rangle
\end{eqnarray*}

\begin{eqnarray}
=\sum\limits_{j^{\prime\prime}m^{\prime\prime}n^{\prime\prime}\lambda}\;
C^{j^{\prime}}_{m^{\prime}}\;^{j}_{m}\;
^{j^{\prime\prime}\lambda^{*}}_{m^{\prime\prime}}\;
C^{j^{\prime}}_{n^{\prime}}\;^{j}_{n}\;^{j^{\prime\prime}\lambda}
_{n^{\prime\prime}}\; D^{j^{\prime\prime}}_{m^{\prime\prime}
n^{\prime\prime}} (g) |g\rangle .
\end{eqnarray}

\noindent Here the $C$'s are the usual Clebsch-Gordan coefficients
for the decomposition of the direct product $D^{j^{\prime}}\times
D^{j}$ of two UIR's into UIR's $D^{j^{\prime\prime}}$, with a
multiplicity index $\lambda$ to keep track of multiple occurrences
of a given $D^{j^{\prime\prime}}$.  If we introduce the shorthand
notation
\begin{eqnarray}
C^{j^{\prime}}_{m^{\prime}n^{\prime}}\;^{j}_{mn}\;^{j^{\prime\prime}}
_{m^{\prime\prime}n^{\prime\prime}} =\sum\limits_{\lambda}\;
C^{j^{\prime}}_{m^{\prime}}\;^{j}_{m}\;^{j^{\prime\prime}\lambda^{*}}
_{m^{\prime\prime}}\;C^{j^{\prime}}_{n^{\prime}}\;^{j}_{n}\;
^{j^{\prime\prime}\lambda}_{n^{\prime\prime}} ,
\end{eqnarray}

\noindent we get from (A.1):
\begin{eqnarray}
U(j^{\prime}m^{\prime}n^{\prime}) U(jmn)
=\sum\limits_{j^{\prime\prime}m^{\prime\prime}n^{\prime\prime}} \;
C^{j^{\prime}}_{m^{\prime}n^{\prime}}\;^{j}_{mn}\;^{j^{\prime\prime}}
_{m^{\prime\prime}n^{\prime\prime}}\;
U(j^{\prime\prime}m^{\prime\prime}n^{\prime\prime}).
\end{eqnarray}

\noindent We can now tackle the product of four factors in
eq.(5.4b).  First using eqs.(3.7,4.18) and then using (A.3) above
gives:
\begin{eqnarray*}
U(j^{\prime}m^{\prime}n^{\prime}) V(g^{\prime}) U(jmn) V(g) =
U(j^{\prime}m^{\prime}n^{\prime}) \sum\limits_{k}\;
D^{j}_{mk}\left(g^{\prime -1}\right) U(jkn) V(g^{\prime} g)
\end{eqnarray*}

\begin{eqnarray}
=\sum\limits_{j^{\prime\prime}m^{\prime\prime}n^{\prime\prime}k}\;
D^{j}_{mk} \left(g^{\prime
-1}\right)\;C^{j^{\prime}}_{m^{\prime}n^{\prime}}\;^{j}_{kn}
\;^{j^{\prime\prime}}_{m^{\prime\prime}n^{\prime\prime}}\;
U(j^{\prime\prime}m^{\prime\prime}n^{\prime\prime})\;V(g^{\prime}g),
\end{eqnarray}

\noindent which is eq.(5.4b).

\end{document}